# PERFORMANCE EVALUATION OF FD-SOI MOSFETS FOR DIFFERENT METAL GATE WORK FUNCTION

Deepesh Ranka[1], Ashwani K. Rana[2], Rakesh Kumar Yadav[3], Kamalesh Yadav[4], Devendra Giri[5]

[#]*Department of Electronics and Communication, National Institute of Technology, Hamirpur Hamirpur (H.P)-177005, India*

[1]ranka.deepesh@gmail.com
[2]ashwani_paper@yahoo.com
[3]rintu2008@gmail.com
[4]kamleshnhr@gmail.com
[5]*devendragiri20@gmail.com*

***ABSTRACT***

*FULLY DEPLETED (FD) SILICON ON INSULATOR (SOI) METAL OXIDE FIELD EFFECT TRANSISTOR (MOSFET) IS THE LEADING CONTENDER FOR SUB 65NM REGIME. THIS PAPER PRESENTS A STUDY OF EFFECTS OF WORK FUNCTIONS OF METAL GATE ON THE PERFORMANCE OF FD-SOI MOSFET. SENTAURUS TCAD SIMULATION TOOL IS USED TO INVESTIGATE THE EFFECT OF WORK FUNCTION OF GATES ON THE PERFORMANCE FD-SOI MOSFET. SPECIFIC CHANNEL LENGTH OF THE DEVICE THAT HAD BEEN CONCENTRATED IS 25NM. FROM SIMULATION WE OBSERVED THAT BY CHANGING THE WORK FUNCTION OF THE METAL GATES OF FD-SOI MOSFET WE CAN CHANGE THE THRESHOLD VOLTAGE. HENCE BY USING THIS TECHNIQUE WE CAN SET THE APPROPRIATE THRESHOLD VOLTAGE OF FD-SOI MOSFET AT SAME VOLTAGE AND WE CAN DECREASE THE LEAKAGE CURRENT, GATE TUNNELLING CURRENT AND SHORT CHANNEL EFFECTS AND INCREASE DRIVE CURRENT.*

## 1. INTRODUCTION

Silicon technologies have progressed faster year to year. The main issue must be concentrate about silicon technologies is effects of reducing the dimension of devices. The scaling down of devices is strongly required to achieve high integration density and better device performance. Due to reduction in the channel length the short channel effects and leakage current become important issue that degrades the device performance in terms of leakage current and short channel effects [1]. To overcome the problem, a new circuit design techniques has been introduce for a newer technologies such as Silicon-on-Insulator (SOI). SOI refers to placing a thin layer of silicon on top of an insulator [2], usually silicon dioxide (SiO2) or known as buried oxide layer (BOX). MOSFETs fabricated on SOI substrate that having a relatively thin SOI layer is known as fully depleted SOI and for thick SOI layer is known as partially depleted SOI. Usually, for fully-depleted SOI devices, the thickness of silicon is about less than bulk depletion width [3]. The full isolation in SOI device provide many advantages such as the drain-to-substrate capacitance can be neglected due to insulator (SiO2) that having dielectric constant lower than Silicon [4]. In recent years, silicon-on insulator (SOI) has attracted considerable attention as a potential alternative substrate for low power application [5]. The most common approach for reducing the power is power supply scaling. Since power supply reduction below three times threshold voltage ($3V_t$) will degrade circuit speed significantly, scaling of the power supply should be accompanied by threshold voltage reduction [6, 7]. However, the lower limit of the threshold voltage is set by the amount of off-state leakage current that can be tolerated, ideally should be no less than 0.4V [6]. Because of BOX layer the fully depleted silicon-on-insulator (FDSOI) has the advantages of lower parasitic junction capacitance and better sub-threshold swing, reduced short channel effects, and free of kink effect [2, 8, 9]. Especially with the geometry scaling down to deep sub-micron range, the threshold voltage also scale down, therefore, threshold voltage control is becoming more important for the future technology.

The SOI MOSFETs also suffer from short channel effects in the sub 65 nm regime due to reduction in threshold voltage [10]. Due to scaling of MOSFET threshold voltage is also continuously decreasing; As a result leakage current and short channel effects also increasing. A metal gate technology can overcome these issues by providing the appropriate gate work functions. Work function is the minimum energy (usually measured in electron volts) needed to remove an electron from a solid to a point immediately outside the solid surface In order to maintain good short-channel performance and proper threshold voltages, the gate work functions of the n-FETs and p-FETs must be close to those of n and p doped poly-Si for bulk-Si CMOS devices and within 0.2 eV of Si mid-gap in novel structures such as SOI MOSFETs. By changing the work function of metal gates of SOI MOSFET we can set the appropriate threshold voltage at same supply voltage and also able to reduce short cannel effects and leakage currents. Previously [17] the gain, transconductance and gate tunnelling current are not considered but in this work these parameters variation with metal gate work function is shown.

However, we can scale the traditional bulk MOSFET device structure down into the 10-nm gate length regime a heavy channel doping will be required to control short channel effects because by using body doping we can reduce the depletion layer width of the MOSFET devices[11]. But result is not comparable to the SOI MOSFET, and also this presents a challenge in terms of device fabrication. However, even if we can achieve this, it will result in the degradation of device performance. Because highly doped channel brings problem with reduced carrier mobility and random dopants fluctuations.

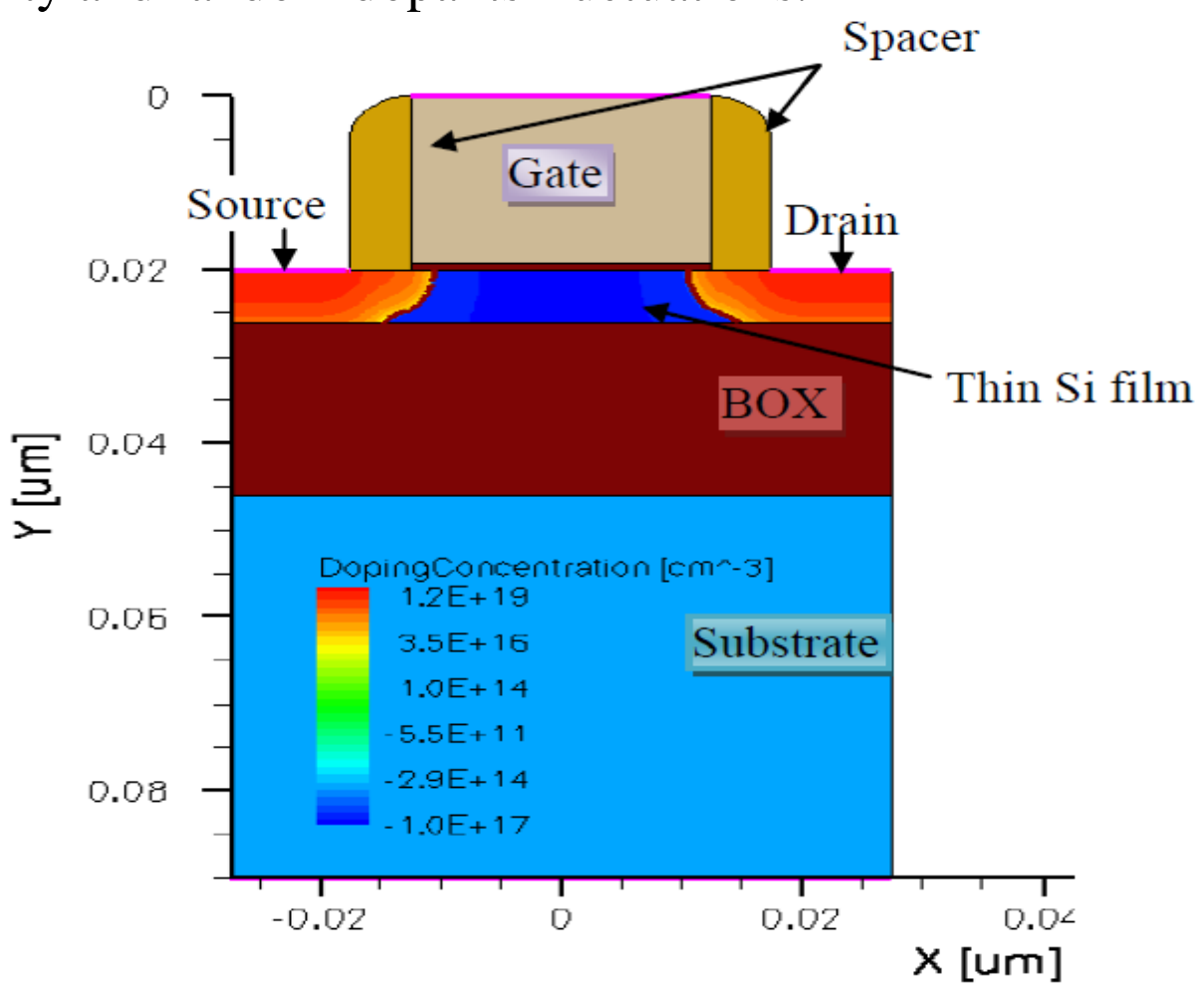

Figure 1. Illustration of simulated structure of FD-SOI NMOSFET [18].

We can also set the appropriate threshold voltage of SOI MOSFET device by providing channel doping and thin Si-film variation but it degrades the device performance in terms of SCE, carrier mobility and dopant fluctuations [12]. A large channel doping will also increase band-to-band tunneling leakage between the body and drain [13]. As a result, both channel doping and Si-film variation result in degraded device performance. This emphasizes the need for gate work function engineering as alternative solutions for nano meter devices .By using Work function engineering we can change the threshold voltage of SOI MOSFET and we are able to maintain device performance and also able to get better result in terms of short channel effects and leakage current. Molybdenum (Mo) is a potential candidate for a metal gate technology provided that a sufficient and stable work function shift can be obtained. The work function of Mo can be significantly reduced by high dose nitrogen implantation. Nitrogen implantation can be used to adjust the Mo gate work function in a controllable way without degrading device performance. Hence by altering the work function of the metal gate of SOI MOSFET, we can set the appropriate threshold voltage, and we are able to reduced short channel effects, gate tunneling current and leakage current.

## 2. Short Channel Effects in FD-SOI MOSFETs

A MOSFET device is considered to be short when the channel length is the same order of magnitude as the depletion-layer widths ($x_{dD}$, $x_{dS}$) of the source and drain junction. As the channel length $L$ is reduced to increase both the operation speed and the number of components per chip, the so-called short-channel effects arise [10].

The short-channel effects are attributed to two physical phenomena:

1. The limitation imposed on electron drift characteristics in the channel
2. The modification of the threshold voltage due to the shortening channel length.

In this letter two short channel effects has been discussed and analyzed:

1. Subthreshold slope(SS)
2. Drain induced barrier lowering (DIBL)

### 2.1. Threshold Voltage of FD-SOI

The analysis provided here is for NMOS, with its extension to PMOS device being straightforward. The threshold voltage of an n-channel MOSFET is classically given [8] by:

$$V_{th} = \phi_{MS} - \frac{Q_{ss}}{C_{ox}} + 2\phi_F + \frac{qN_a x_{d\max}}{C_{ox}} \tag{1}$$

Where $\phi_{MS}$ is the work function difference between the gate and the channel and equal to $\phi_m - (\phi_{Si} - \phi_F)$. $Q_{ss}$ is the surface state charge of the channel. $C_{ox}$ is the gate capacitance and equal to $\frac{\varepsilon_{ox}\varepsilon_o}{t_{ox}}$. $t_{ox}$ is the gate oxide thickness. $\phi_F$ is the Fermi potential, equal to $\frac{kT}{q}\ln(\frac{N_a}{n_i})$, where $N_a$ is the channel doping concentration. $x_{d\max}$ is the maximum depletion width, for partially depleted device equal to $\sqrt{\frac{4\varepsilon_{si}\phi_F}{qN_a}}$, Equation (1) works well for NMOS device.

For the fully depleted silicon on insulator devices, the silicon film is very thin compared to the partially depleted devices. Under the thin silicon film, there is a very thick buried oxide. Similar to the gate oxide, there are some surface charges between the silicon film and the buried oxide. This surface charge varies with the technology. When the silicon film scales down below 100nm, these charges have effects to the threshold voltage. In this paper we model this effect as a capacitor. For the fully depleted devices is equal to the silicon film thickness $t_{Si}$. Then the threshold voltage for the fully depleted silicon on insulator devices changes to:

$$V_{th} = \phi_{MS} - \frac{Q_{ss} t_{ox}}{\varepsilon_{ox}\varepsilon_o} + 2\phi_F + \frac{qN_a t_{Si} t_{ox}}{\varepsilon_{ox}\varepsilon_o} - Q_{ssb}\left(\frac{t_{ox}}{\varepsilon_{ox}\varepsilon_o} + \frac{t_{Si}}{\varepsilon_{Si}\varepsilon_o}\right) \quad (2)$$

Where the $Q_{ssb}$ is the surface charge between the silicon film and the buried oxide, $t_{Si}$ is the silicon film thickness

### 2.2. Subthreshold Slope

It indicates how effectively the flow of drain current of a device can be stopped when $V_{gs}$ is decreased below $V_{th}$. When Id-Vg curve of a device is steeper subthreshold slope will improve.

Subthreshold slope [11] is given by:

$$S_t = \left[\frac{d(\log_{10} I_{ds})}{dV_{gs}}\right]^{-1} = \frac{kT}{q}\left(1 + \frac{C_d}{C_i}\right) \quad (3)$$

Where $C_d$ = depletion layer capacitance

$C_i$ = gate oxide capacitance

A device characterized by steep subthreshold slope exhibits a faster transition between off (low current) and on (high current) states.

### 2.3. Drain Induced Barrier Lowering (DIBL)

In the weak inversion regime, there is a potential barrier between the source and the channel region. The height of this barrier is a result of the balance between drift and diffusion current between these two regions. The barrier height for channel carriers should ideally be controlled by the gate voltage to maximize transconductance. The DIBL effect [12] occurs when the barrier height for channel carriers at the edge of the source reduces due to the influence of drain electric field, upon application of a high drain voltage. This increases the number of carriers injected into the channel from the source leading to an increased drain off-current. Thus, the drain current is controlled not only by the gate voltage, but also by the drain voltage.

For device modelling purposes, this parasitic effect can be accounted for by a threshold voltage reduction depending on the drain voltage [13].

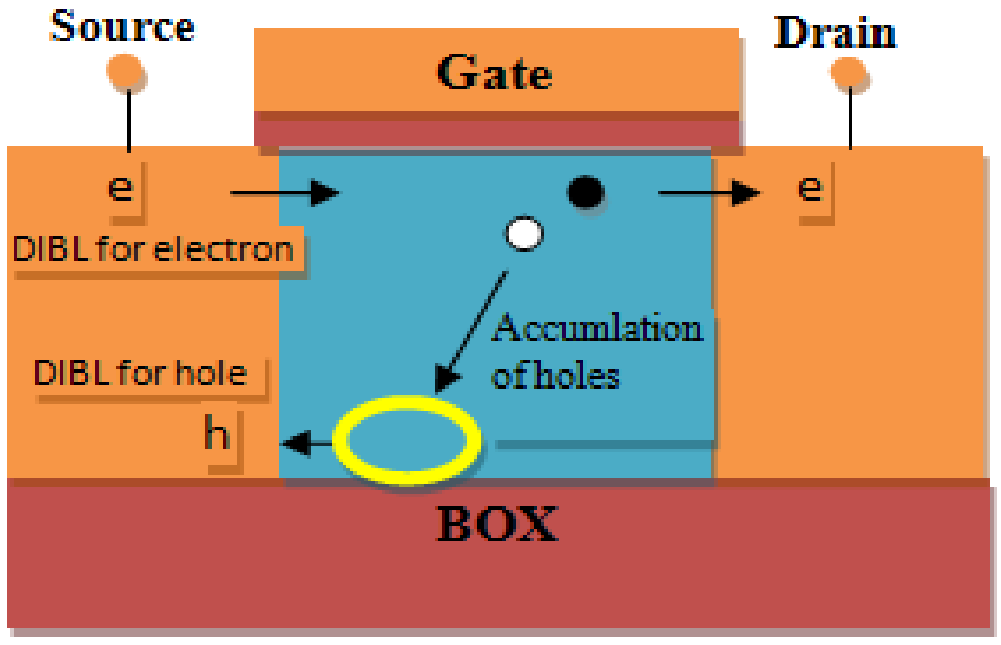

Figure 2. Three mechanisms determining SCE in SOI MOSFETs [14].

In addition to the surface DIBL, there are two unique features determining SCEs in thin-film SOI devices: 1) positive bias effect to the body due to the accumulation of holes generated by impact ionization near the drain and 2) the DIBL effect on the barrier height for holes at the edge of the source near the bottom of thin film, as illustrated in figure 2 [14].

# 3. Transport Description

The Sentaurus Device input file simulates the $I_dV_g$ performance of an SOI NMOSFET device using the basic drift-diffusion transport model.

## 3.1. Drift diffusion model

The conduction in this model is governed by the Poisson's equation (4) and the equations of continuities of the carriers (5), (6). The Poisson's equation which couples the electrostatic potential $V$ to the density of charge is given by

$$\nabla^2 V = \frac{-q}{\varepsilon}\left[p - n + N_D^+ + N_A^- + n_T\right] \quad (4)$$

Where $n$ and $p$ represent the densities of the electrons and the holes, respectively, $N_D^+$ and $N_A^-$ are the ionized donor and acceptor impurity concentrations, respectively, $n_T$ is the density of carriers due to the centre of recombination [17] and $\varepsilon$ is the dielectric constant. The transport equations express the current densities of the electrons and the holes; they are composed of two components, drift and diffusion [17].

$$\vec{J}_n = q\mu_n n \vec{E} + qD_n \vec{\nabla}_n , \quad (5)$$

$$\vec{J}_p = q\mu_q p - qD_p \vec{\nabla}_p \quad (6)$$

Where $\mu_n$ and $\mu_p$ are the electron and hole mobilities, $D_n$ and $D_p$ are the diffusion coefficients of electrons and holes and $\vec{\nabla}_n$ and $\vec{\nabla}_p$ are the two-dimensional gradients of concentration of electrons and holes. $\vec{E}$ is the electric field applied. The equations of continuities represent the carriers conservation in a volume element for the electrons and the holes, respectively.

$$\frac{\partial n}{\partial t} = GR_n + \frac{1}{q}\vec{\nabla} J_n \quad (7)$$

$$\frac{\partial p}{\partial t} = GR_p - \frac{1}{q}\vec{\nabla} J_p \quad (8)$$

The $GR_n$ and $GR_p$ terms describe the phenomena of recombination-generation and $J_n$ and $J_p$ are the current densities [17]. In a steady state regime, electro concentrations $n$, holes $p$, electrostatic potential $V$, and electrical current $I$ are obtained from the solutions of eqs (4)-(8) using the finite differences method. The solution of the equations consists of the discretization of the field in a finite number of points and the approach of the partial derivative, using finite differences method, in all the interior nodes of the field while taking in account the boundary conditions in order to obtain a linear system of equations in the following matrix form:

$$[M].[X] = [b],$$

Where $M$ is the total matrix and $b$ is the vector source.

To solve this system, we use the iterative method of Gausse-Seidel. It is particularly well adapted to the matrices of this type because they do not require additional storage and also present a good numerical convergence.

## 4. Device Structure

To study the work function engineering effect on SOI MOSFET a schematic cross-sectional view of the SOI MOSFET is simulated using 2-D Sentaurus device simulator [18], is shown in figure 1. We assumed light channel doping concentration ($1\times10^{-17}$ $cm^{-3}$) to avoid degrading of carrier mobility and more $V_t$ variations. The doping concentration of source/drain region is kept at $1\times10^{-19}cm^{-3}$. Gate length of the device that had been concentrated is 25nm. Silicon film thickness, gate oxide (SiO2) thickness and BOX thickness are 6nm, 0.6nm and 20nm respectively. Spacer is of Si3N4 and 0.7nm thick. We assumed n channel device and simulated the device for different work function of metal gates of SOI MOSFET. We assumed Mo metal for metal gates because it provides wide range of work function. Metal gate technology may potentially replace conventional poly-Si gate technology for CMOS devices for better performance**.** Molybdenum (Mo) has very low resistivity and high melting point ,and thin films of Mo with (110) crystallographic texture have been shown to exhibit work functions close to 5eV on several candidate dielectrics. It has also been reported that the work function of Mo can be significantly reduced by high-dose nitrogen implantation. Nitrogen implantation can be used to adjust the Mo gate work function in a controllable way without degrading transistor performance. So by using this technique we are able to set the appropriate threshold voltage for maintaining device performance compare to the body doping.

## 5. Simulation Results

In this section different performance parameter of FD-SOI MOSFET for different values of metal gate work function has been discussed. Drift diffusion model has been used for extracting different parameters. All parameters have been extracted at 25nm technology node.

### 5.1. Transconductance $g_m$:

It measures the drain current variation with a gate-source voltage variation while keeping the drain-source voltage constant and is of crucial importance because it decides the ability of the device to drive a load. The transconductance has a important role in determining the switching speed of a circuit and voltage gain of MOSFET amplifiers. High transconductance devices yield circuits capable of high speed operation.

Transconductance of a MOSFET

$$g_m = \left.\frac{\Delta I_{ds}}{\Delta V_{gs}}\right|_{V_{ds}} \qquad (9)$$

From equation (9) transconductance measured by the slope of $I_{ds}$-$V_{gs}$ curve. When Id-Vg curve of a device is steeper it shows better transconductance.

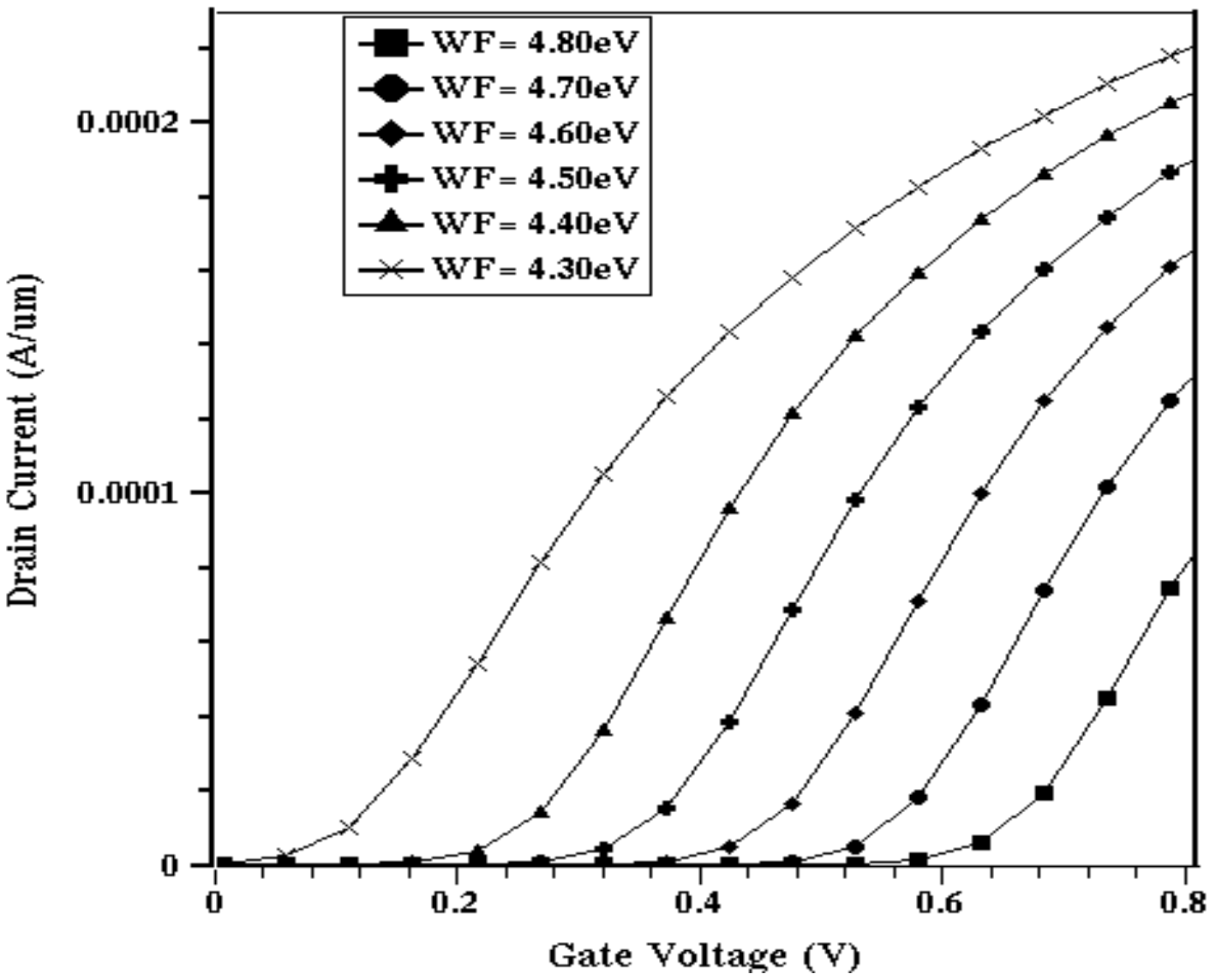

Figure 3. Drain current as a function of gate voltage for different metal gate work function.

Figure 3 shows the drain current as a function of gate voltage for different metal gate work function [19]. This curve actually indicates the transconductance of SOI MOSFETs at different work function of metal gate. As we decrease work function of metal gate transconductance of device increase. This is because from equation (2) as metal gate work function reduce threshold voltage reduces means at low gate voltage channel formed so drain current increase further transconductance.

Voltage gain of a MOSFET device

$$A_v = g_m R_D \tag{10}$$

From equation (10) low metal gate work function MOSFET also provides high voltage gain.

## 5.2. Output Conductance $g_d$:

It measures the drain current variation with a drain-source voltage variation while keeping the gate-source voltage constant.
It is a vital parameter for a device because it decides the drive current of a device.

Output conductance of a MOSFET

$$g_d = \left.\frac{\Delta I_{ds}}{\Delta V_{ds}}\right|_{V_{gs}} \tag{11}$$

Figure 4 shows the variation of drain current as a function of drain voltage. As metal gate work function reduces drain current increase so output conductance also increases. This is because from equation (2) as metal gate work function reduce threshold voltage reduces means at low gate voltage channel formed so drain current increase.

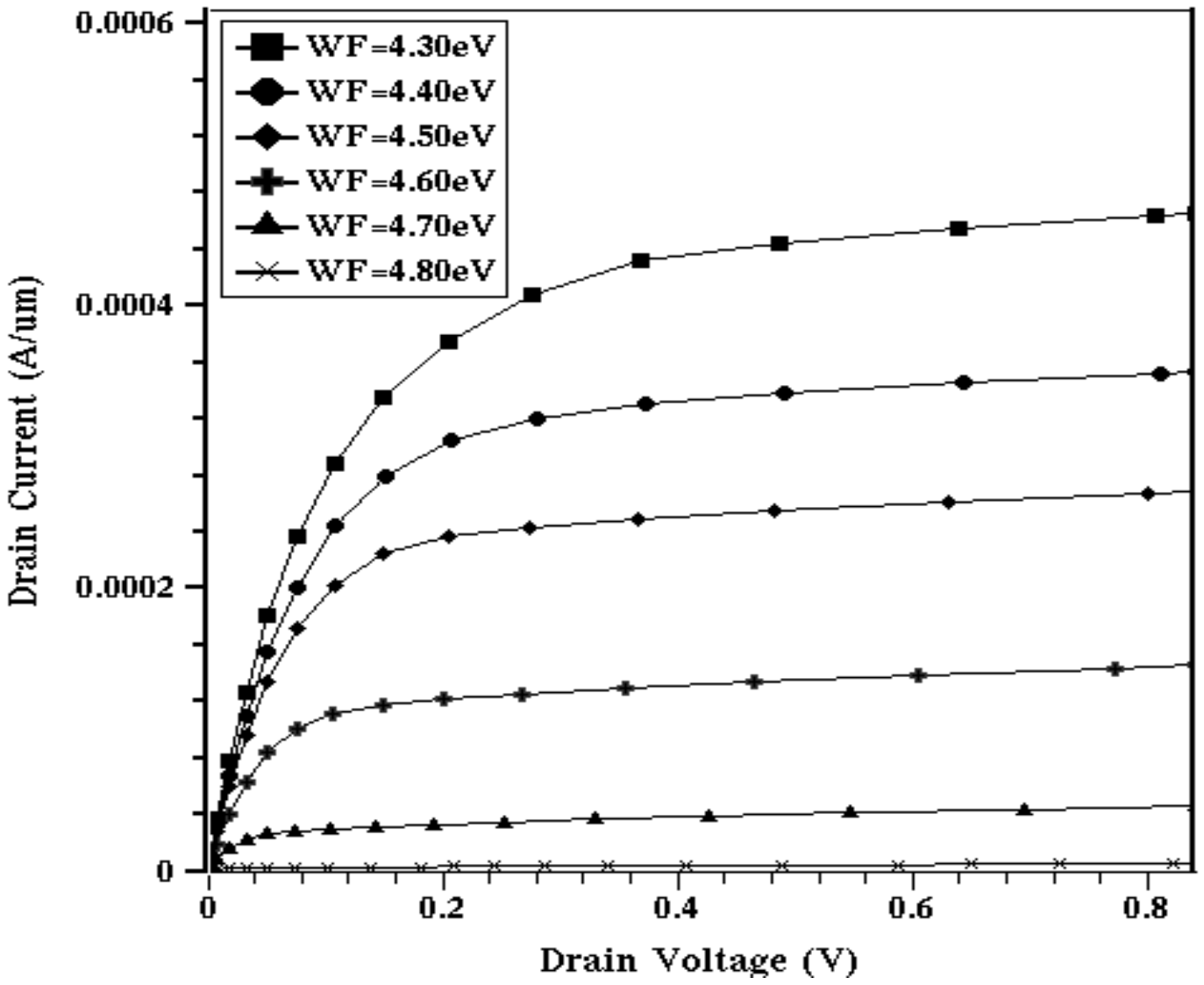

Figure 4. Drain current as a function of drain voltage for different metal gate work function.

### 5.3. Gate Tunnelling Current:

Figure 5 shows the gate leakage current variation with gate voltage at different work function metal gate. As we increase the metal gate work function gate leakage current decreases. From figure 5 it also examines that when we increase gate voltage gate tunnelling current increases. But after 1v it is saturated to a constant value.

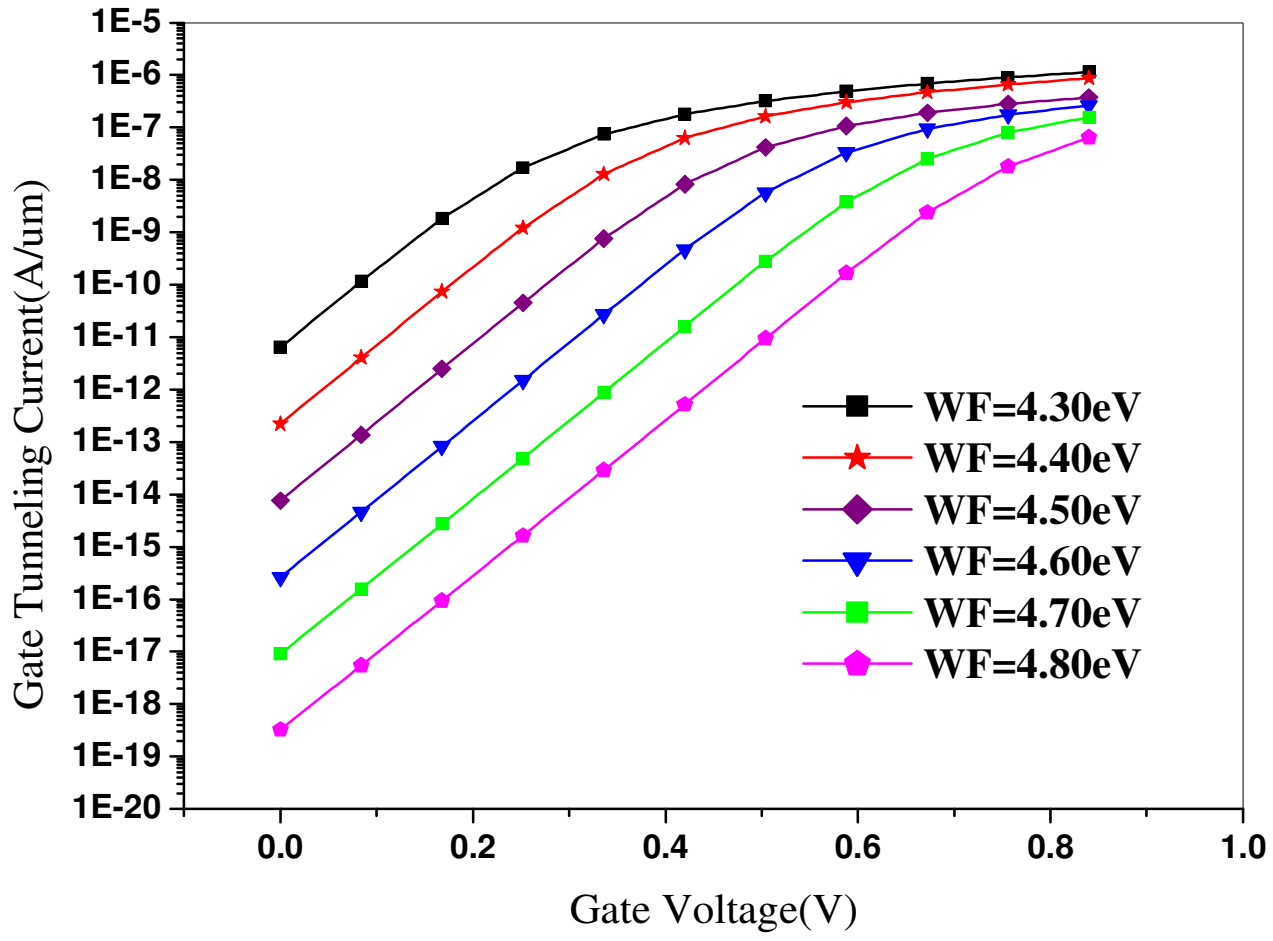

Figure 5. Gate tunnelling current Vs gate voltage for different metal gate work function.

### 5.4. Electron Concentration:

Figure 6 shows the variation of electron concentration along the channel direction for different metal gate work function at particular 1v gate potential. Work function is the minimum energy needed to remove an electron from a solid to a point immediately outside the solid surface. So for low values of work function electron concentration is higher at the channel.

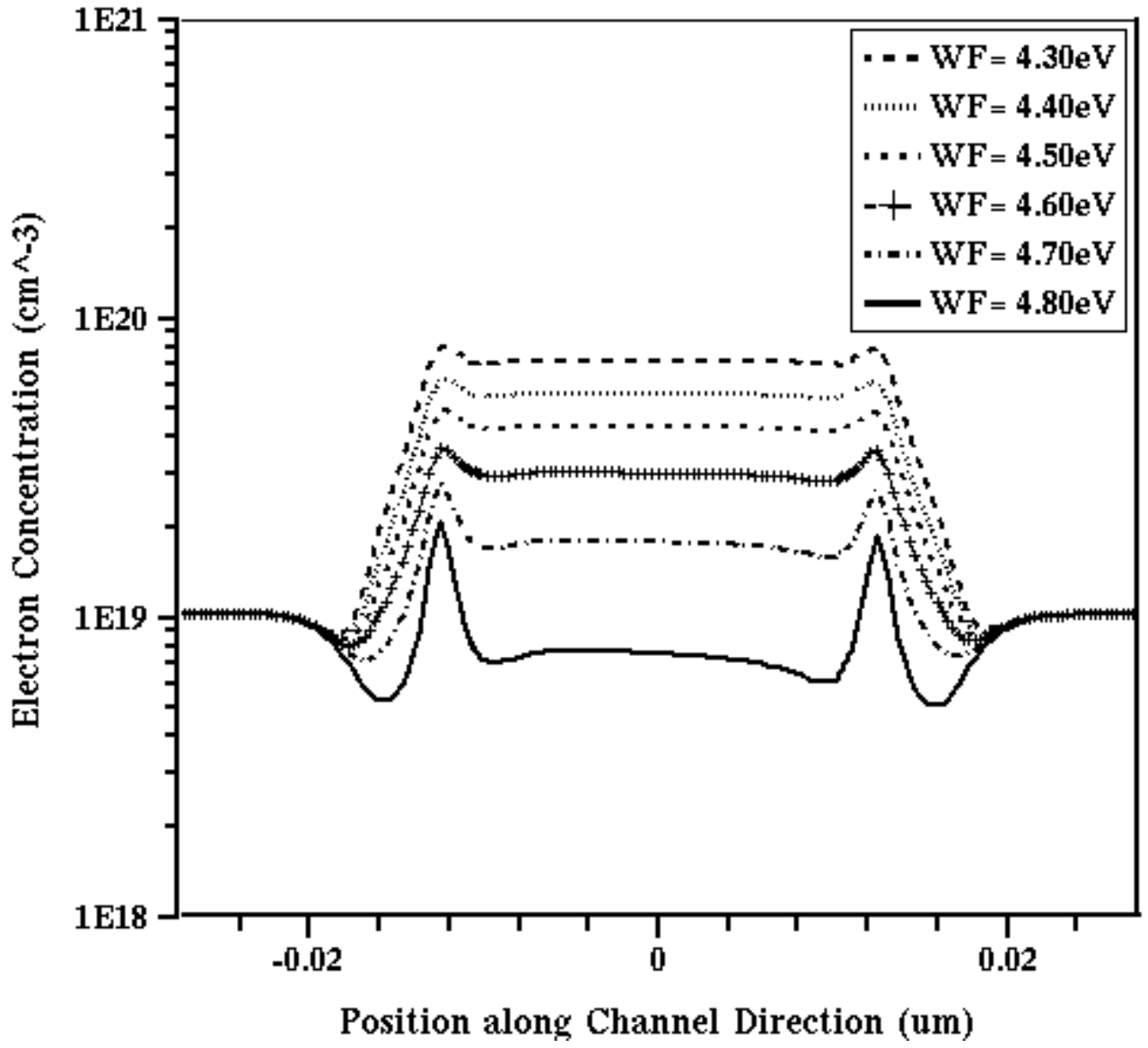

Figure 6. Electron concentration Vs position along channel direction for different metal gate work function at 1v gate potential.

### 5.5. Vertical Electric Field:

Figure 7 shows the variation of electron concentration along the channel direction for different metal gate work function at particular 1v gate potential. From curve it evaluated that for lower values of metal gate work function the vertical electric field along the channel direction is high.

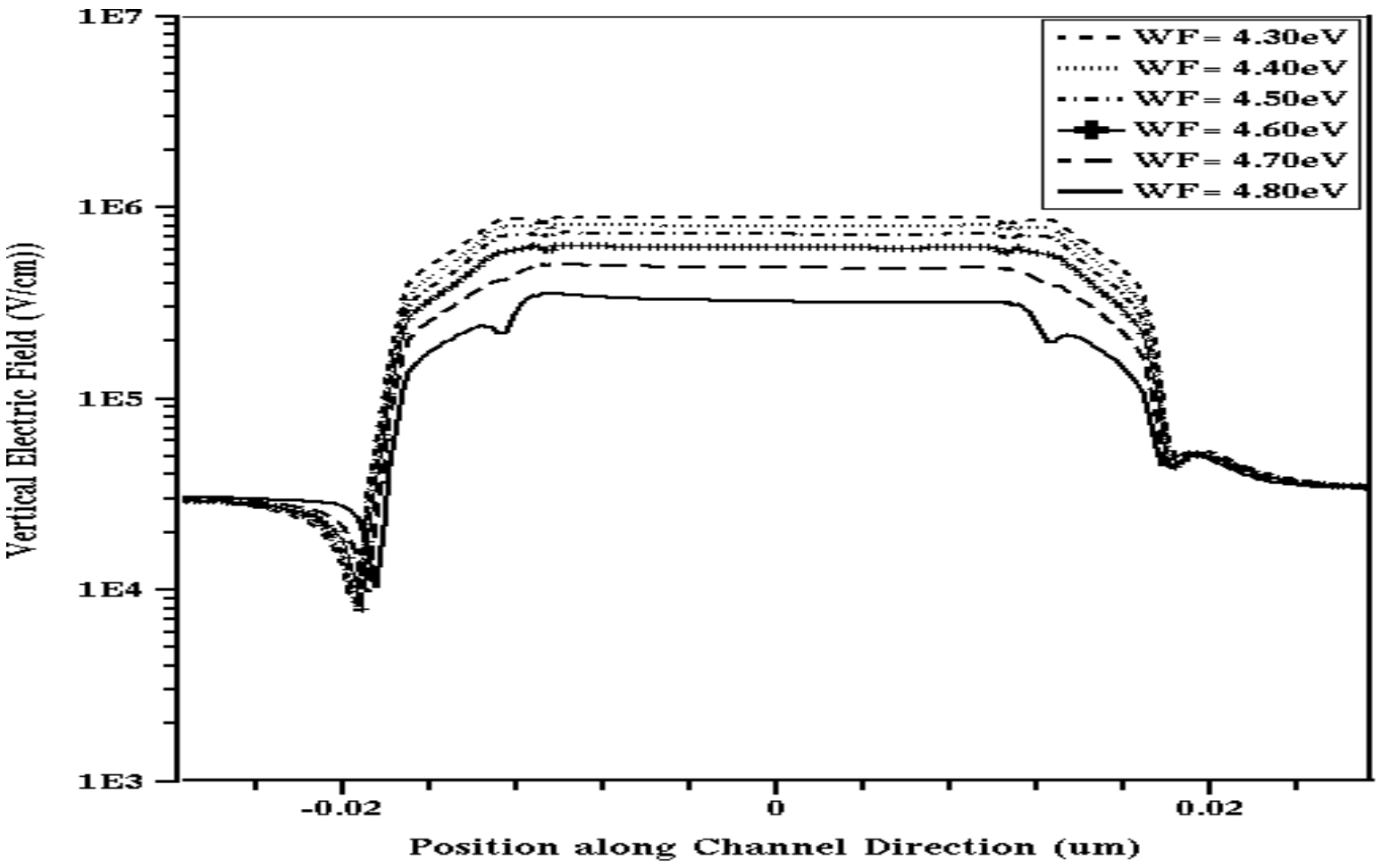

Figure 7. Vertical electric field Vs position along channel direction for different metal gate work function at 1v gate potential.

### 5.6. Threshold Voltage Variation:

From simulation results figure 8 shows that the threshold voltage of SOI MOSFETs is increasing linearly with the increasing of metal gate work function. It is also shown in equation (2) that $V_{th}$ is dependent on the $\phi_{MS}$ (work function difference of metal and semiconductor). It is because higher $V_{gs}$ is required to invert the channel compare to other devices. So by providing the different work function of metal gates of SOI MOSFET device we can set the appropriate threshold voltage.

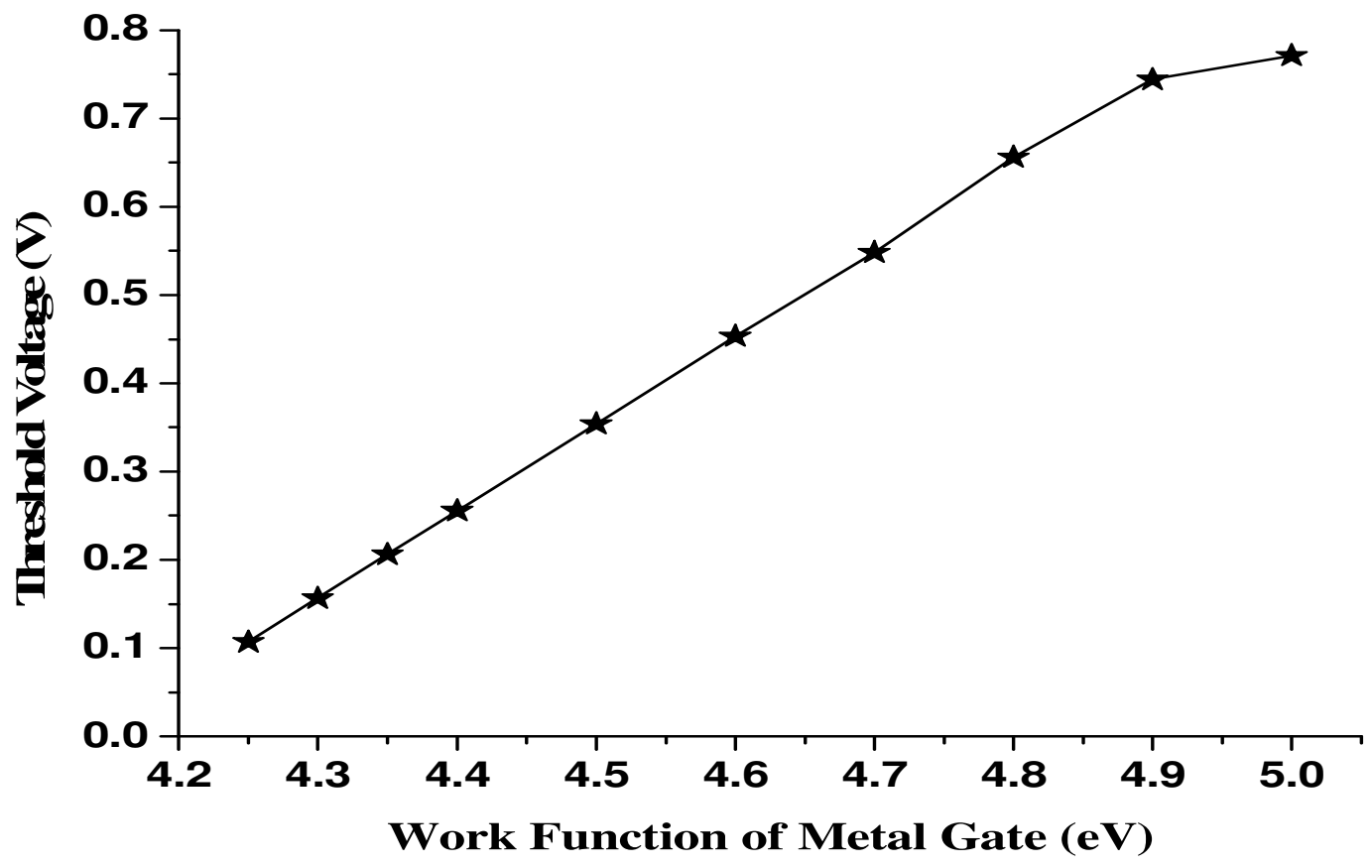

Figure 8. Threshold Voltage variation with different metal gate work function.

### 5.7. Leakage Current (OFF Current) $I_{OFF}$:

Figure 9 shows the variation in leakage current as function of work function of metal gate. As the work function of metal gate increases threshold voltage increase, the leakage current of SOI MOSFET decreases exponentially with threshold voltage.

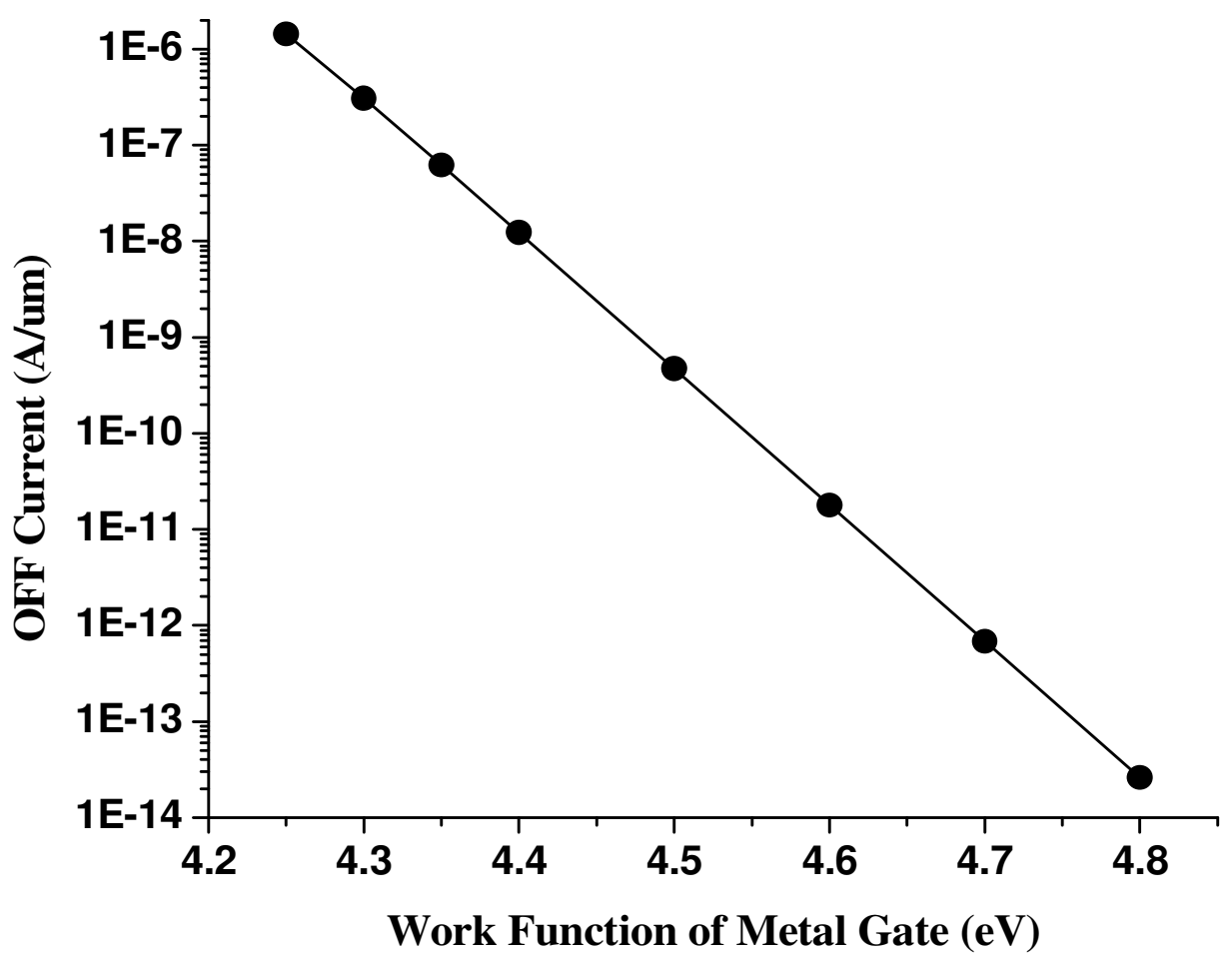

Figure 9. Leakage current analysis at different metal gate work function

### 5.8. On Current OFF Current ratio ($I_{ON}/I_{OFF}$):

Device has higher value of on current to off current ratio ($I_{on}/I_{off}$) provide high switching speed. Figure 10 shows the variation in $I_{on}/I_{off}$ with different metal gate work function. It seems from Fig. that as we increase the work function of metal gate $I_{on}/I_{off}$ increases.

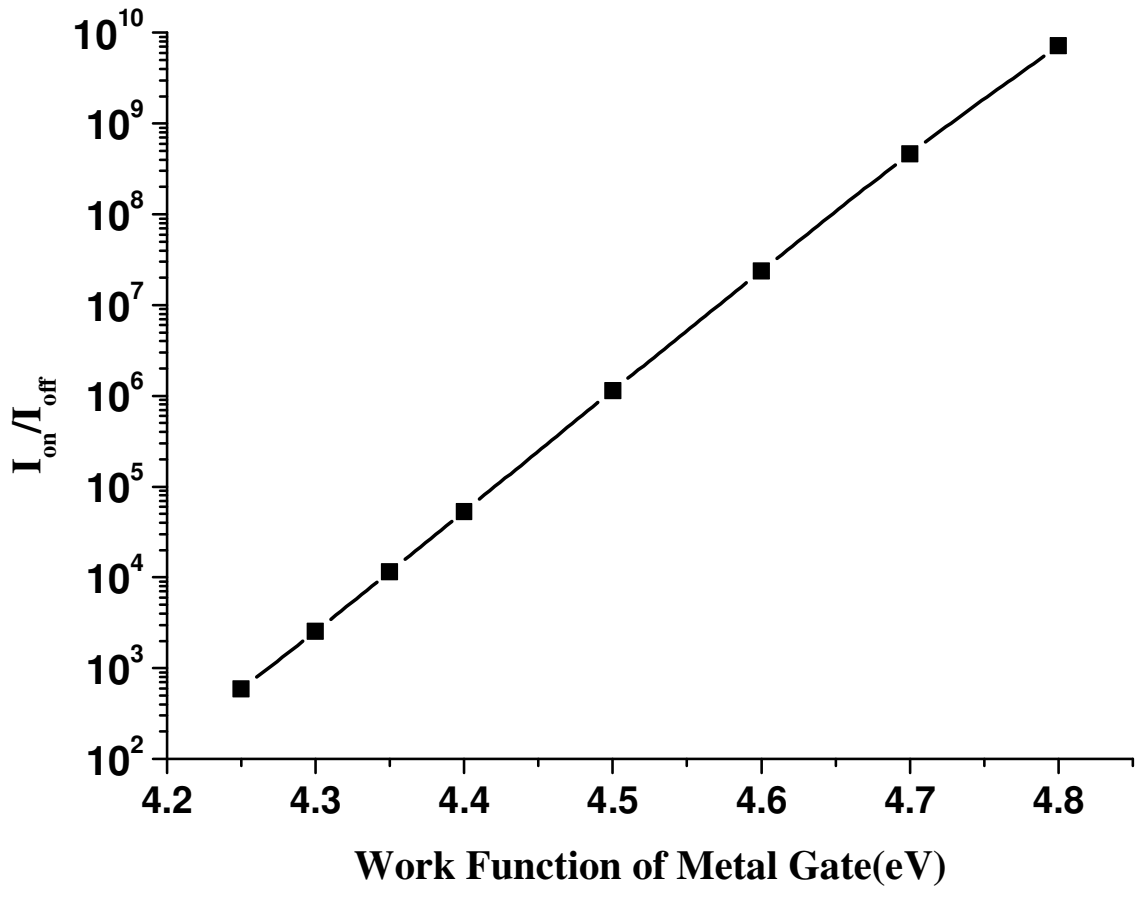

Figure 10. $I_{on}/I_{off}$ variation with metal gate work function.

### 5.9. Selection of Appropriate Metal Gate Work Function for FD-SOI MOSFET

In order to maintain $I_{off}$ very low, it is necessary to increase the metal work function. In addition, the rise in metal work function is accompanied by an increase in threshold voltage. For this sake, the choice of the metal work function must be a delicate compromise between electric performance (reduction of the $I_{off}$ current) and commutation rate (shift from the blocked state to an active state) associated to $V_{th}$. Figure 11 shows the evolution of the $I_{off}$ current and the threshold voltage $V_{th}$ for different metal work function.

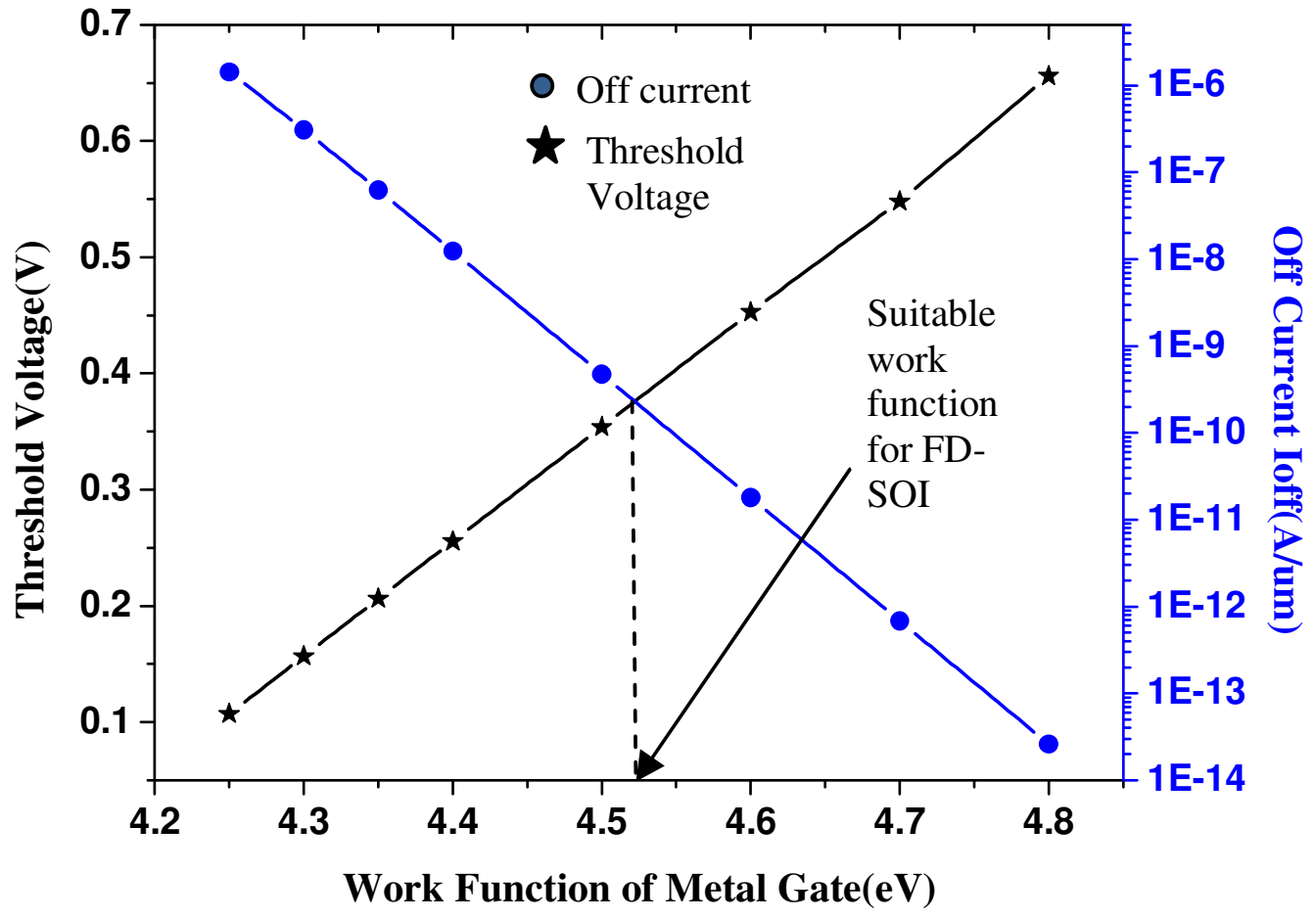

Figure 11. Threshold voltage and leakage current variation for different metal gate work function

So combining figure 8 and figure 9 we find from figure 11 that around 4.50eV metal gate work function optimum value of threshold voltage and leakage current is achieved [17]. Also from figure 3,4,5,10 we see that at 4.50eV metal gate work function it provides better values of transconductance, $I_{dmax}$, $I_{on}/I_{off}$ ,gate tunnelling current respectively.

From figure 12 it shows that the impact of metal gate work function on drain-induced barrier lowering (DIBL) and subthreshold slope (S) are negligible, but for channel lengths LCH less than 10 nm the effect becomes important. As the work function increases subthreshold slope decreases at the same time DIBL increases so we choose particular value of work function for which both DIBL and subthreshold slope have better values.

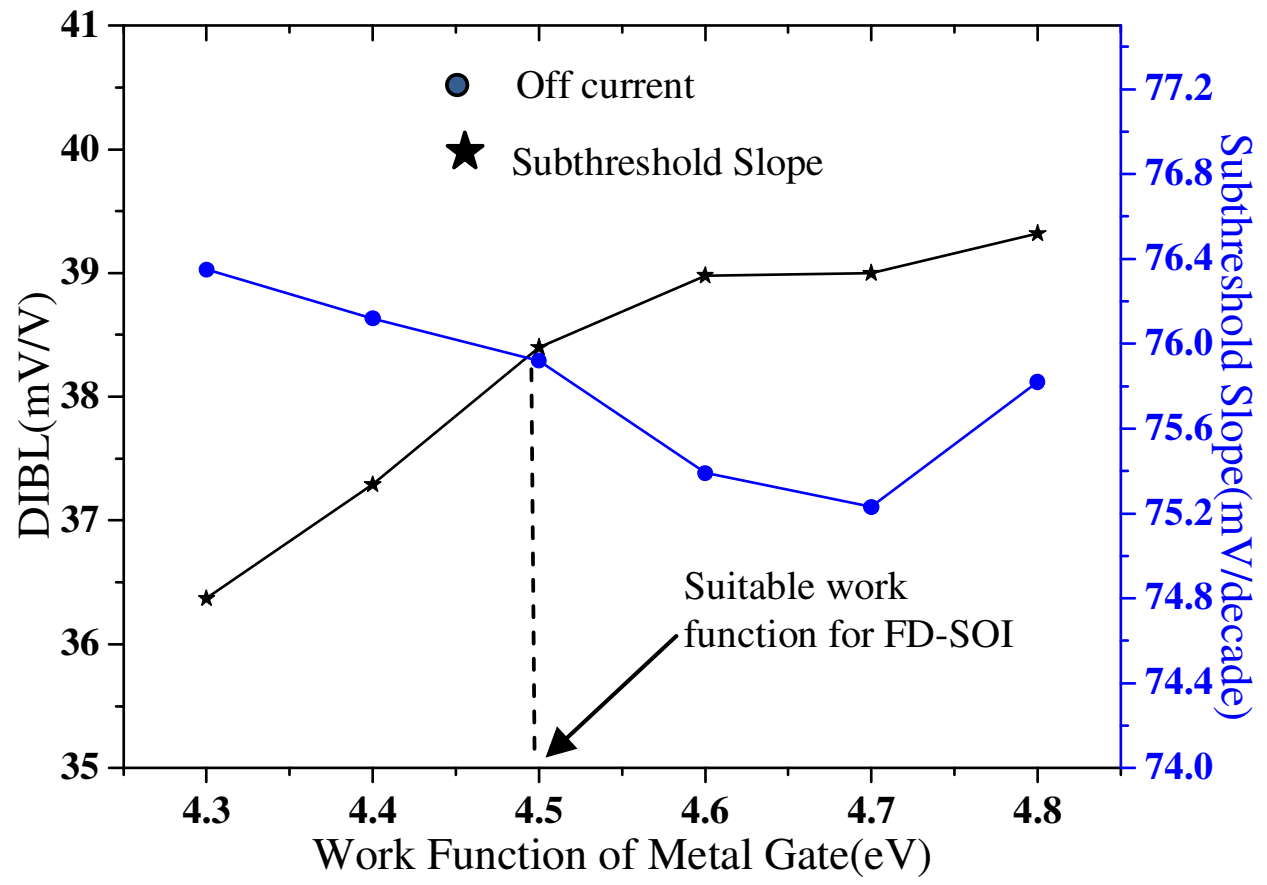

Figure 12. Threshold voltage and leakage current variation for different metal gate work function

## 6. CONCLUSION

Continuous down scaling of MOSFET device is required to increase the device speed and packaging density, but it degrades the device performance in terms of short channel effect and leakage current . For continue scaling down there is need of device structure that provide better performance in deep submicron regime. Due to reduction in the channel length scaling, threshold voltage is decreasing that increasing the leakage current and short channel effects. Channel doping and Si-film thickness variation are the concept by which we can set the desired threshold voltage, of SOI MOSFET device but it degrades the device performance in terms of carrier mobility and dopant fluctuations. The work function engineering is the concepts by which we can set the appropriate threshold voltage of SOI MOSFET device. Variation in channel doping and Si-film thickness doesn't provide linear variation in threshold voltage. But variation in metal gate work function provides linear variation in threshold voltage. So work function engineering is preferred over other two methods because this method provides better device performance and linear variation in threshold voltage. As we increase the work function threshold voltage increases and leakage current decreases also DIBL increases and subthreshold slope decreases so for this sake we choose a particular value of work function at which device provides better performance. For thin body CMOS devices the range of gate work is 4.4eV to 5eV that can be easily done with metallic gates. From simulation it conclude that at 4.50eV metal gate work function SOI MOSFET device shows better performance than any other device. In addition, metallic gates provide immunity from the depletion which allows a thinner effective dielectric thickness without affecting device performance. Molybdenum is an attractive candidate for gate electrode application because of its compatibility with CMOS processing, and its high work function (5eV) makes it ideal as a gate material for lightly doped SOI MOSFETs. The work function of the Mo can be altered by nitrogen implantation without degrading the device performance. Hence, we can set the appropriate threshold voltage by changing the work function of metal gates of SOI MOSFET.